\documentclass[useAMS,usenatbib]{mn2e}

\usepackage{graphicx}
\usepackage{subfigure}
\usepackage{color} 
\usepackage{amsmath}
\def\lsim{\lower.5ex\hbox{$\; \buildrel < \over \sim \;$}}
\def\gsim{\lower.5ex\hbox{$\; \buildrel > \over \sim \;$}}
\def\simeq{\lower.3ex\hbox{$\; \buildrel \sim \over - \;$}}
\def\ch{\lower-0.55ex\hbox{--}\kern-0.55em{\lower0.15ex\hbox{$h$}}}
\def\lh{\lower-0.55ex\hbox{--}\kern-0.55em{\lower0.15ex\hbox{$\lambda$}}}
\def\eg{{\it e.g.,} }
\def\etal{{\em et al.} }
\def\ie{{\em i.e.,} }
\title[Periodic massloss from viscous accretion flows]
{Periodic massloss from viscous accretion flows around black holes}
\author[Santabrata Das, Indranil Chattopadhyay, Anuj Nandi, Diego Molteni]
{Santabrata Das$^{1}$\thanks{E-mail:
sbdas@iitg.ernet.in (SD); indra@aries.res.in (IC); anuj@isac.gov.in (AN);
diego.molteni@unipa.it (DM)}, 
Indranil Chattopadhyay$^{2}$, 
Anuj Nandi$^{3}$,
D. Molteni$^{4}$
\footnotemark[1] \\ 
$^{1}$Indian Institute of Technology Guwahati, Guwahati, 781039, India. \\
$^{2}$ARIES, Manora Peak, Naintal, 263002, India.\\
$^{3}$Space Astronomy Group, SSIF/ISITE Campus, ISRO Satellite Centre, Outer  
Ring Road, Marathahalli, Bangalore, 560037, India. \\
$^{4}$Dipartimento di Fisica e Chimica, University of Palermo, Italy.}
\begin{document}

\date{Accepted . Received ; in original form }

\pagerange{\pageref{firstpage}--\pageref{lastpage}} \pubyear{}

\maketitle

\label{firstpage}

\begin{abstract}

We investigate the behaviour of low angular momentum viscous accretion
flows around black holes using Smooth Particle Hydrodynamics (SPH)
method. 
Earlier,
it has been observed that in a significant part of the energy and angular momentum parameter space,
rotating transonic accretion flow
undergoes shock transition before entering in to the black hole and a 
part of the post-shock matter is ejected as bipolar outflows,
which are supposed to be the precursor of relativistic jets. In this work,
we simulate accretion flows having injection parameters from the 
inviscid shock parameter space, and study the response of viscosity on them.
With the 
increase of viscosity, shock becomes time dependent and starts to oscillate when 
the viscosity parameter crosses its critical value. As a result, the in falling matter
inside the post-shock region exhibits quasi-periodic variations 
and causes periodic ejection of matter from the inner disc as outflows.
In addition, the same hot and dense post-shock matter emits high
energy radiation and the emanating photon flux also modulates
quasi-periodically. Assuming a ten solar mass black hole, the 
corresponding power density spectrum peaks at
the fundamental frequency of few Hz followed by multiple harmonics.
This feature is very common in several outbursting black hole 
candidates. We discuss the implications of such periodic variations.
\end{abstract}


\section{Introduction}

Luminosities and spectra emanating from the microquasars
and AGNs are best explained by the gravitational energy released due to accretion onto compact objects
such as black holes.
However, it has been established in recent years, that black hole candidates,
be it stellar mass or super massive, emits powerful collimated
outflows or jets \citep{l99}. Since black holes do not have intrinsic atmosphere or hard surface,
these jets have to be generated from the accretion disc itself.
In a very interesting paper, \citet{jetal99}
had shown that jets originate from around a region less
than $100r_g$ ($r_g{\equiv}$ Schwarzschild radius) across the
central object of the nearby active galaxy M87. Since the typical timescale
of an AGN or a microquasar scales with mass, temporal behaviour of black hole
candidates is studied with microquasars \citep{mkkf06}.  After investigating the connection
between accretion and ejection in ten microquasars, \citet{gfp03} 
concluded that mildly relativistic, quasi steady jets are generally ejected in the low hard
spectral states (\ie when electromagnetic spectra peak in the high energy power-law frequency range)
of the accretion disc. It was also shown that jets tend to get stronger
as the microquasar moves to the intermediate hard states, and truly relativistic ejections
are observed during hard-intermediate to soft-intermediate transition, after which the microquasar enters
canonical high soft state (\ie when spectra peak in the thermal low energy range),
which shows no jet activity \citep{rsfp10,mil12}. All these evidences
suggest that the generation or quenching of jets do
depend on various states of the accretion disc, and that, 
the jet formation mechanism is linked with the processes
that dominate at distances relatively closer to the black hole. 

It is well known that, spectra from microquasar change states between, the low hard state (LH)
and high soft state (HS), through a series of
intermediate states. Interestingly, the hard power-law photons exhibit a quasi-periodic oscillations (QPO).
The QPOs evolve along with the spectral states of the accretion disc,
starting with low frequencies in LH, increasing
as the luminosity increases, and reaches highest
value before disappearing in the HS state \citep{cdnp08,st09,ndmc12,nrs13}. 
Interestingly, 
although QPO frequency increases as the accretion disc moves
from LH to intermediate states, but no QPO was detected during ejection of relativistic jets \citep{nrs13,rn13} 
which suggests that, probably the part of the disc responsible for QPO is entirely ejected as relativistic jets.
This conversely also suggests that, the inner part of the disc is responsible for QPOs and is also the base of the
jet.
Accretion disc models which are invoked to explain the accretion-ejection phenomena around black hole candidates,
should at least address the connection between spectral states, QPO evolution and the evolution of jets, apart from matching
the luminosities radiated by AGNs and microquasars.

There are various accretion disc models in the literature. We know that, matter
crosses the horizon with the speed of light ($c$) and circular orbits cannot exist within the marginally stable orbit
($3r_g$).
So the inner boundary condition for accretion onto
black hole is necessarily transonic, as well as,
sub-Keplerian, which implies that advection should be significant at least close to the horizon.
The very first model of accretion onto black holes was of course radial
inflow of matter, which was basically the general relativistic version of the Bondi
solutions \citep{b52,m72}. However, the infall time scale of radial accretion onto black holes is short,
and therefore has very little time to produce the huge luminosities observed from AGNs and microquasars \citep{s73}.
On the other hand,
\citet{ss73} considered a geometrically thin but optically thick accretion disc characterized by negligible advection,
but by virtue of possessing
Keplerian angular momentum distribution, the disc was rotation dominated.
This disc model was radiatively efficient and produced the multi-coloured blackbody part of the spectra
or the `blue bump' radiated by the AGNs. However, the presence of hard power-law tail in the spectra of the black hole
candidates indicated the necessity of a hot Comptonizing cloud, which was neither present in Keplerian disc
nor its origin could be identified in any self-consistent manner from such a disc.
Therefore, models with advection gained importance. Theoretically, it
was shown that in a significant
range of energy and angular momentum, multiple sonic points may exist for rotating,
transonic accretion flows onto black holes, where the existence of the inner sonic point
is purely due to the presence of gravity stronger than that due to the Newtonian variety \citep{lt80}.
It has been shown
numerically as well as analytically, that such transonic matter
in the multiple sonic point regime, may undergo steady or
non-steady shock transition. Shock in accretion may be pressure
and centrifugally supported, if the flow is rotating
\citep{f87,c90,mlc94,mrc96,msc96,cd04,mgt06,cd07,d07,dc08} or only be
pressure supported if the flow is spherical \citep{co85,ke86,bom89}.
The most popular amongst accretion disc models with advection is the so-called
advection dominated accretion flow (ADAF), and it is characterized by a single sonic point
close to the horizon \citep{nkh97}. It has been shown later,
that ADAF type solution is a subset of a general viscous advective
accretion solutions \citep{lgy99,bdl08,kc13}.
The shock in accretion disc around black holes has been shown to exist for 
multispecies flows with
variable adiabatic index ($\gamma$) as well \citep{c08,cc11,kscc13,ck13}.

Shock transition for accretion flow are favourable mechanism to
explain many of the observational features of black hole candidates.
Hot electrons in the post-shock region, abbreviated as CENBOL
(CENtrifugal pressure supported Boundary Layer),
may explain
the power-law tail of the spectrum from black hole candidates in
hard states, while a weak or no shock solution produces a dearth of hot electrons which may cause
the weaker power-law tail in the soft states \citep{ct95,mc10}. Moreover, a
large number of authors have shown the formation of bipolar outflows from the
post-shock accretion flow, both numerically \citep{mlc94,msc96} as
well as analytically \citep{lb05,cd07,fk07,dc08,dbl09,kc13,kscc13,kcm14}. It is also
interesting to note that, by considering a simplified inviscid accretion,
and which has the right parameters to form a standing shock,
\citet{dcnc01} qualitatively showed that there would be no jets in no-shock or weak shock
condition of the disc, or in other words, when the disc is in the soft spectral state.
This indicates the conclusions of
\citet{gfp03}. Such a scheme of accretion-ejection solution is interesting because, the jet
base is not the entire accretion disc but the inner part of the disc, as has been
suggested by observations \citep{jetal99,detal12}.

Although, most of the efforts have been undertaken theoretically to
study steady shocks, perhaps transient shock formations may explain
the transient events of the black hole candidates much better.
\citet{msc96}, considered bremsstrahlung cooling of an inviscid
flow, and reported there is significant oscillation of the
post-shock region. Since the post-shock flow is of higher density
and temperature compared to the pre-shock flow, the cooling rates
are higher. If the cooling timescale roughly matches with the
infall timescale at the shock the resonance condition occurs and
the post-shock flow oscillates. Since the post-shock region is
the source of hard X-rays \citep{ct95}, thus its oscillation
would be reflected in the oscillation of the emitted hard
X-rays --- a plausible explanation for QPOs \citep{cm00}.
In this paper, we will focus on the
oscillation of the shock front, but now due to viscosity instead
of any cooling mechanism. \citet{cd04,kc13} had shown that, with the
increase of viscosity parameter, in the energy-angular momentum parameter
space, the domain of shock decreases.  
We know viscosity transports angular momentum outwards, while the
specific energy increases inwards. How does the general flow
properties of matter, which are being launched with same injection
speed and temperature, be affected with the increase in viscosity
parameter? It has been shown from simulations and theoretical studies
that post-shock matter is ejected as jets, however if the shock is weak
then the jet should be of lower power! We would like to find the condition of the shocked disc
that produces weak or strong jets. Disc instability due to viscous transport
has been shown before and has been identified with QPOs \citep{lmc98,lcscbz08,lrc11},
however, we would like
to show how this instability might affect the shock induced bipolar outflows.
Moreover, it has been shown theoretically that the energy and angular momentum for which
steady shock exists in inviscid flow, will become unstable for viscous flow \citep{cd04,kc13}.
We would like to see how the mass outflow rate depend on unstable shock,
or in other words, if there is any connection between QPOs and mass outflow rate.
In this paper, we would like to address these issues.

In the next section, we present the governing equations and model
assumptions. In section 3, we present the results, and in the
last section we draw concluding remarks.

\section{Governing equations}

We consider a non-steady accretion flow around a non-rotating
black hole. The space time geometry around a Schwarzschild black
hole is modeled using the pseudo-Newtonian potential introduced
by \citet{pw80}.

In this work, we use geometric units as $2G=M_B= c = 1$,
where $G$, $M_B$ and $c$ are the gravitational constant, the mass of
the black hole and the speed of light, respectively. In this unit
system, distance, velocity and time are measured in units of
$r_g=2GM_B/c^2$, $c$ and $t_g=2GM_B/c^3$ respectively, and the equations have been
made dimensionless.

The Lagrangian formulation of the two-dimensional fluid dynamics equations
for SPH \citep{m92} in cylindrical coordinate are given by \citep{lmc98} --- \\
The mass conservation equation,
$$
\frac{D \rho}{D t} = -\rho \nabla.{\bmath v},
\eqno(1)
$$
where, $\frac{D}{Dt}$ denotes the co-moving time-derivative and $\rho$ is
the density.

The radial momentum equation is given by,
$$
\frac{D v_r}{Dt} = -\frac{1}{\rho} \frac{\partial P}
{\partial r} + g_r + \frac{v^2_{\phi}}{r}.
\eqno(4a)
$$

The vertical momentum equation is,
$$
\frac{Dv_z}{Dt} = -\frac{1}{\rho}\frac{\partial P}
{\partial z} + g_z.
\eqno(4b)
$$

The azimuthal momentum equation is,
$$
\frac{D v_{\phi}}{Dt} = -\frac{v_{\phi} v_r}{r}
+\frac{1}{\rho} \left[ \frac{1}{r^2}\frac{\partial}
{\partial r} \left( r^2 \tau_{r\phi} \right) \right],
\eqno(4c)
$$
where, $\tau_{r \phi}$ is the $r-\phi$ component of viscous stress tensor
and is given by,
$$
\tau_{r\phi}=\eta r \frac{\partial \Omega}{\partial r},
\eqno(4d)
$$
and the angular velocity is given by,
$$
\Omega = \frac{v_{\phi}}{r}.
\eqno(4e)
$$

In Eqs. 4(a-b), $g_r$ and $g_z$ are the components of gravitational force
\citep{pw80} and are given by,
$$
g_r = - \frac{1}{2(R-1)^2}\frac{r}{R}, \hskip 1.0cm {\rm and}
\hskip 1.0cm
g_z = - \frac{1}{2(R-1)^2}\frac{z}{R},
\eqno(4f)
$$
where $R = \sqrt{r^2 + z^2} $.

The form of dynamic viscosity parameter is given by \citep{ss73},
$$
\eta = \nu \rho = \alpha \rho a h 
$$
where, $\nu$ is the kinematic viscosity, $\alpha$ is the viscosity parameter,
 $a~(=\sqrt{{\gamma P}/{\rho}})$ is the sound speed, 
$\rho$ is the mass density, $h~[= \sqrt{\frac{2}{\gamma}}ar^{1/2}(r-1)]$ is the disc
half height estimated using hydrostatic equilibrium \citep{cd04}
and $v=\sqrt{v^2_r+v^2_z}$. 

The conservation of energy is given by,
$$
\frac D{Dt}\left( \epsilon +\frac 12{\bf v}^2\right) =-\frac P\rho \nabla
{\bf v+v\cdot }\left( \frac{D{\bf v}}{Dt}\right) +\frac 1\rho \nabla \left(
\bf \vec{ {\bf \tau } } {\bf :v} \right),
\eqno(4g)
$$
with 
$$
\left( \frac{D{\bf v}}{Dt}\right) =-\frac 1\rho \nabla P+{\bf g}, 
$$

\noindent where, $P=\left( \gamma -1\right) \rho \epsilon$ is the
equation of state of ideal gas, g is the gravitational
acceleration and $\epsilon$ is the internal energy, respectively.
${\bf \vec{{\bf \tau }}:v}$ is the vector resulting from the
contraction of the stress tensor with the velocity vector.
We include only $\tau _{r\phi }$ (namely, the $r-\phi$ component)
since it is the dominant contributor to the viscous stress.
A complete steady solution requires the equations of energy,
angular momentum and mass conservation supplied by transonic
conditions at the critical points and the Rankine-Hugoniot conditions
at the shock.

\section{Results}

\begin{figure}
\begin{center}
\includegraphics[width=0.5\textwidth]{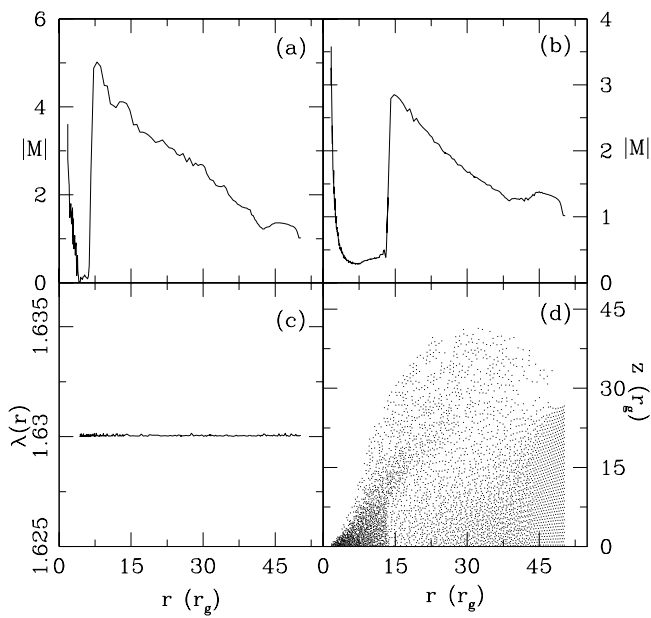}
\end{center}
\caption{(a) Mach number $M=v_r/a$ on the equatorial plane with $r$ at $t=317.39~t_g$, (b) $M$ on the equatorial plane
with $r$, after the
solution has reached steady state. (c) Specific angular momentum $\lambda$ on the equatorial plane with $r$,
and (d) the distribution of SPH particles in steady state in $r-z$ plane. The injection parameters
are injection velocity $v_{inj} = -0.06436$, sound speed $a_{\rm inj} = 0.06328$
and angular momentum $\lambda_{\rm inj} = 1.63$ at $r_{inj}=50.4$. See text for details.}
\end{figure}

\begin{figure}
\begin{center}
\includegraphics[width=0.5\textwidth]{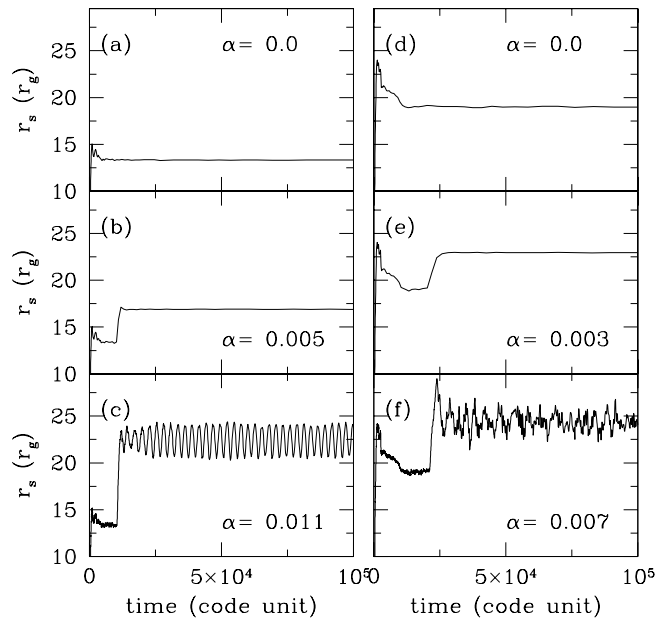}
\end{center}
\caption{
Time evolution of shock location. Viscosity
parameter chosen are (a) $\alpha = 0.0$ (b) $0.005$ and (c) $0.011$,
for injection parameters $r_{\rm inj} = 50.4~r_g$ with
injection velocity $v_{\rm inj} = -0.06436$, sound speed $a_{\rm inj} = 0.06328$
and angular momentum $\lambda_{\rm inj} = 1.63$. And the panels on the right
are plotted for viscosity (d) $\alpha=0$, (e) $0.003$ and (f) $0.007$,
for injection parameters $r_{\rm inj} = 50.8~r_g$ with
injection velocity $v_{\rm inj} = -0.06532$, sound speed $a_{\rm inj} = 0.06221$
and angular momentum $\lambda_{\rm inj} = 1.7$.
Here, the adiabatic index $\gamma = 4/3$.}
\end{figure}

In order to obtain the time dependent axisymmetric, viscous
accretion solution, we adopt Smooth Particle Hydrodynamics
scheme. Here, we inject SPH
particles with radial velocity $v_{\rm inj}$, angular momentum
${\lambda}_{\rm inj}$ and sound speed $a_{\rm inj}$ at the injection
radius $r_{\rm inj}$. Initially, the accreting matter is treated as
inviscid in nature. At the injection radius, the disc height is
estimated considering the fact that the flow remain in hydrostatic
equilibrium in the vertical direction and obtained as 
$H_{\rm inj} \sim a_{\rm inj} r^{1/2}_{\rm inj}(r_{\rm inj}-1)$.
With the suitable choice of flow parameters at the injection radius,
accretion flow
may undergo shock transition. For a given set of flow parameters like the Bernoulli parameter
${\cal E}=0.00449$ and the specific angular momentum
$\lambda=1.63$, we plot the equatorial Mach number $M=v_r/a$
of the flow with $r$ at $t=317.39t_g$, and the transient shock is at $r_s=7.5r_g$ (Fig. 1a).
The steady state is reached at
$t=10^4t_g$, and the stationary shock settles at $r_s=15r_g$ (Fig. 1b). 
The steady state angular momentum distribution
on the equatorial plane is shown in Fig. 1c.
The position of the SPH particles in steady state
is shown in Fig. 1d in the $r-z$ plane.

Once the steady state is achieved in the inviscid limit, we turn on the
viscosity. 
It must be pointed out that turning on the viscosity after obtaining
the inviscid steady state shock solution, 
doesn't affect our conclusions
since exactly the same result would be obtained if the viscosity is turned on initially.
However, since turning on $\alpha$ makes the numerical code slow, it would have taken much
longer time to search the parameters for which the disc admits shock solution.
The role of viscosity is to remove angular momentum outwards,
and consequently it perturbs the shock front, and may render the stationary shock unstable.
In Fig. 2a, we show
the time evolution of the shock location 
for inviscid accretion flow. The shock location is measured at the disc equatorial plane.
Here, we use input parameters as $r_{\rm inj} = 50.4$,
$v_{\rm inj} = - 0.06436$, $a_{\rm inj} = 0.06328$, and 
${\lambda}_{\rm inj} = 1.63$ respectively.
For $\alpha = 0.005$, we find stable shock at around $17r_g$
depicted in Fig. 2b. When viscosity is increased further and reached
its critical limit, namely $\alpha = 0.011$,
shock front starts oscillating and the oscillation sustains forever,
provided the injected flow variables remain unaltered. This feature is
shown in Fig. 2c. 
For further increase of viscosity, the oscillation
becomes irregular and for even higher $\alpha$, the shock oscillation
is irrevocably unstable.
We change the injection parameters to $v_{\rm inj} = - 0.06532$, $a_{\rm inj} = 0.06221$, and 
${\lambda}_{\rm inj} = 1.7$ at $r_{\rm inj} = 50.8$, and plot the time evolution
of the shock $r_s$ for inviscid flow (Fig. 2d), $\alpha=0.003$ (Fig. 2e) and $\alpha=0.007$
(Fig. 2f). The mechanism of shock oscillation due to the presence of viscosity, may be 
understood in the following manner. We know viscosity transports angular momentum ($\lambda$) outwards.
Since the post-shock disc is hot, the angular momentum transport in the post-shock disc
is more efficient than the pre-shock disc.
Accordingly, angular momentum piles up in the post shock region. 
On the other hand, a steady shock forms if the momentum flux, energy flux and mass flux are conserved
across the shock. Therefore, as the angular momentum piles up in the immediate post-shock
region, extra centrifugal force will try to push the shock front outward. If this piling up of $\lambda$
is moderate then the expanded shock front will find equilibrium at some position to form steady shock
(\eg Figs. 2b, 2e).
If this outward push is strong enough then the expanding shock front will overshoot a possible
equilibrium position and in that case the shock front will oscillate (\eg Figs. 2c, 2f).  
If the angular momentum piling results in too strong centrifugal barrier,
it could drive the shock out of the computation domain.

It may be noted that we are simulating the inner part of the disc, \ie the inner few $\times~10r_g$ of the disc.
And, therefore the injection parameters are similar to the inner boundary conditions of an accretion disc.
Since matter entering a black hole has angular momentum, typically $1 \lsim \lambda \lsim 2$, our chosen angular momentum
are also of the same order. However, at the outer edge angular momentum may reach very high value depending
on $\alpha$. For example, with injection parameters $r_{\rm inj} = 50.4$,
$v_{\rm inj} = - 0.06436$, $a_{\rm inj} = 0.06328$, and 
${\lambda}_{\rm inj} = 1.63$, one can integrate the effective one dimensional equations of motion 
to estimate the angular momentum at the outer boundary and it is typically
around $45~r_gc$ at $r_{\rm out} \sim 4000~r_g$.

\begin{figure*}
\vbox to220mm{\vfil 
\includegraphics[width=0.9\textwidth]{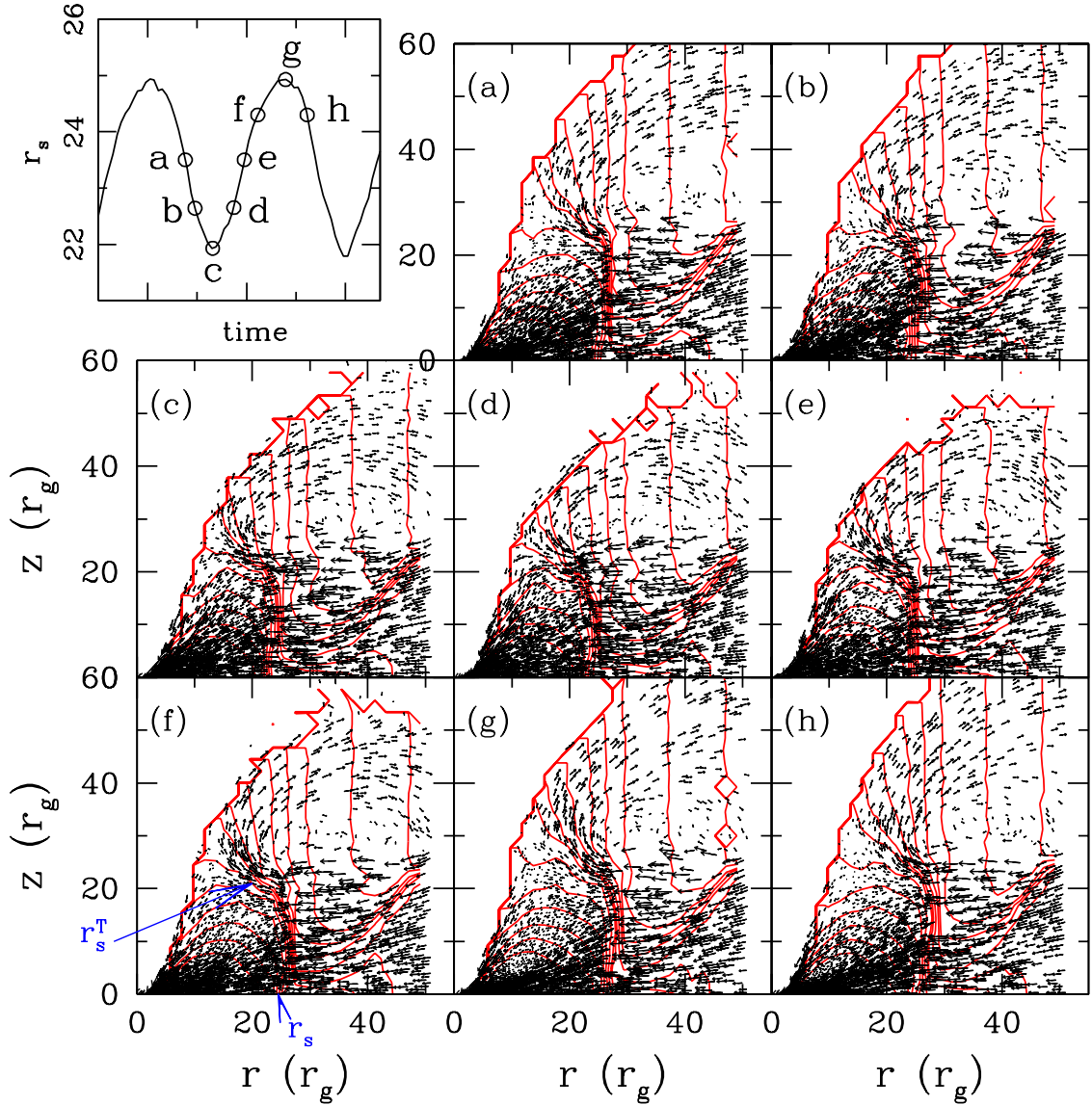}
\caption{Velocity vectors are in the $r-z$ plane. SPH particles having
angular momentum $\lambda_{inj} = 1.63$ are injected supersonically
from  $r_{\rm inj} = 50.4$ with injection velocity $v_{\rm inj} = -0.06436$
and sound speed $a_{inj} = 0.06328$, respectively. Viscosity parameter
is chosen as $\alpha = 0.011$ and adiabatic index $\gamma = 4/3$.
Shock location oscillates with time. Figure 3(a-h) represent the
various snap shots of velocity distribution taken at equal interval
within a complete period of shock oscillation. In Figure 3f, $r^T_s$ and $r_s$
are indicated, and clearly there is a phase lag between the two. Here, Figure 3c denotes
the case when shock location is closest from the black hole and
Figure 3g represents the case when shock location is at its maximum value.
Density contours (solid curves, red online) are
plotted over the velocity vector field.}
\vfil}
\label{landfig}
\end{figure*}

From Figs. 2c-2e, it is clear that the shock can experience persistent oscillation
for some critical viscosity and injection parameters. But the shock front is a surface
and that too not a rigid one. Therefore, every part of the shock front will not
oscillate in the same phase, resulting in a phase lag between the shock front
on and around the equatorial plane and the top of the shock front ($r^T_s$).
For simplicity, we
record the variation of shock location with time at the disc
equatorial plane which is shown in {\it top-left} panel of Fig. 3.
Note that there exists quasi-periodicity in the variation of shock
location with time. We identify eight shock locations within a given 
oscillation period that are marked in open circles. The respective
velocity field and density contours (solid, online red) of the flow in the $r-z$ plane is shown
in the rest of the panels of Figs. 3(a-h).
Higher density and extra thermal gradient force in the CENBOL 
region causes a fraction of in falling matter to bounce-off as
outflow. When shock front oscillates, post-shock volume also
oscillates which induces a periodic variation of driving force
responsible to vertically remove a part of the in falling matter.
As the shock reaches its minimum
(Fig. 3c), the thermal driving of outflow is weak, so the spewed up matter falls back.
The post-shock outflow continues to fall as the shock expands to its maxima
(Figs. 3d, e, f). However, as the shock reaches its maximum value the thermal
driving also recovers (Fig. 3g). The extra thermal driving plus the squeeze of
the shock front as it shrinks, spews strong outflow (Fig. 3g).
In Fig. 3f, we have indicated the shock
location on the equatorial plane $r_s$, and the top of the shock front $r^T_s$.
The position of the shock front can easily be identified from the clustering of the density
contours connecting $r_s$ and $r^T_s$. From Figs. 3(a-h), it is clear that the mass outflow rate is significant 
when $r^T_s \gsim r_s$. 

Due to shock transition, the post shock matter becomes hot and dense
which would essentially be responsible to emit high energy radiation.
At the critical viscosity, since the shock front exhibits regular
oscillation, the inner part of the disc, \ie CENBOL, also
oscillates indicating the variation of photon flux emanating from
the disc. Thus, a correlation between the variation of shock front
and emitted radiation seem to be viable. Usually, 
the bremsstrahlung emission is estimated as,
$$
E_{Brem} = \int_{r_{1}}^{r_{2}} \rho^2 T^{1/2} r^2 dr,
$$
where, $r_1$ and $r_2$ are the radii of interest within which radiation
is being computed and $T$ is the local temperature. In this work,
we calculate the total bremsstrahlung
emission for the matter from the CENBOL region. 
Also, we quantify the mass outflow rate calculated
assuming an annular cylinder at the injection radius which is
concentric with the vertical axis. The thickness of the cylinder
is considered to be twice the size of a SPH particle. This ensures
at least one SPH particle lies within the cylindrical annulus.
We identify particles leaving the computational domain as outflow
provided they have positive resultant velocity, \ie$v_r > 0$
and $v_z > 0$ and they lie above the disc height at the injection
radius. With this, we estimate the mass outflow rate which is defined
as $R_{\dot m}={\rm outflow~rate}~({\dot M}_{\rm out})/
{\rm accretion~rate}~(|{\dot M}_{\rm in}|)$ and observe its
time evolution. 
Here, ${\dot M}_{in(out)} = 2 \pi \rho_{\rm inj(out)}
v_{\rm inj(out)} x_{\rm inj(out)}H_{\rm inj(out)}$.
In Figure 4, we present the variation of shock 
location, corresponding bremsstrahlung flux from the post-shock region and mass
outflow rate with time. Here, the radiative flux is plotted in arbitrary 
unit. Assuming a $10 M_{\sun}$ black hole, the overall time evolution of five seconds
($\equiv$ 50600 code time) is shown for representation.
The input parameters are $r_{\rm inj} = 50.4$, $v_{\rm inj} = -0.06436$,
$a_{\rm inj} = 0.06328$, $\lambda_{\rm inj} = 1.63$ and $\alpha = 0.011$
respectively. Note that persistent shock oscillation takes place
over a large time interval, with the oscillation amplitude $\sim 3 r_g$.
This phenomenon exhibits the emission of non-steady radiative flux
which is nicely accounted as quasi-periodic variation commonly seen
in many black hole candidates \citep{cm00,rm06}. 
Subsequently, periodic mass ejection
also results from the vicinity of the gravitating objects as a 
consequence of the modulation of the inner part of the disc
due to shock oscillation.

To understand the correlation between the shock oscillation
and the emitted photon flux from the
inner part of the disc, we calculate the Fourier spectra of the
quasi-periodic variation of the shock front and the power spectra
of bremsstrahlung flux for matter resides within the boundary of
post-shock region as well as outflow with resultant velocity $v>0$. The obtained
results are shown in Figure 5, where the {\it top panel} is
for shock oscillation, {\it middle panel} is for photon flux variation
from post-shock disc and {\it bottom panel} is for photon flux
variation of outflowing matter, respectively. Here, the input
parameters are same as Figure 4.
We find that the quasi-periodic variation of the
shock location and the photon fluxes from post-shock disc and
outflow are characterized by the fundamental
frequency $\nu_{\rm fund} = 3.7$ Hz which is followed by multiple
harmonics. The first few prominent harmonic frequencies are 
$7.4$ Hz ($\sim 2 \times \nu_{\rm fund}$), $11.2$ Hz
($\sim 3 \times \nu_{\rm fund}$) and $14.2$ Hz
($\sim 4 \times \nu_{\rm fund}$). This suggests that the dynamics
of the inner part of the disc \ie the post-shock disc and emitted fluxes
are tightly coupled. In order to understand the generic nature
of the above findings, we carried out another simulation with 
different input parameters. The results are shown in Figure 6, 
where we use $r_{\rm inj} = 50.4~r_g$, $v_{\rm inj} = -0.06436$,
$a_{\rm inj} = 0.06328$, $\lambda_{\rm inj} = 1.61$ and $\alpha = 0.013$,
respectively. The solutions are obtained similar to the previous case,
\ie first a steady state inviscid solution is obtained and then the viscosity is turned on.
The corresponding Fourier spectra of shock oscillation
and power spectra of radiative
fluxes are presented in the {\it top}, {\it middle} and {\it bottom
panel}, respectively. The obtained frequencies for quasi-periodic
variations are $2.9~{\rm Hz}~{\rm \bf }~(\bf \nu_{\rm \bf fund})$, 
$5.6$ Hz ($\sim 2 \times \nu_{\rm fund}$), $9.3$ Hz
($\sim 3 \times \nu_{\rm fund}$) and $15$ Hz
($\sim 5 \times \nu_{\rm fund}$). 
In both the cases, the obtained power density spectra (PDS)
of emitted radiation has significant similarity with number of
observational results \citep{rm06,ndmc12}.

\begin{figure}
\begin{center}
\includegraphics[width=0.45\textwidth]{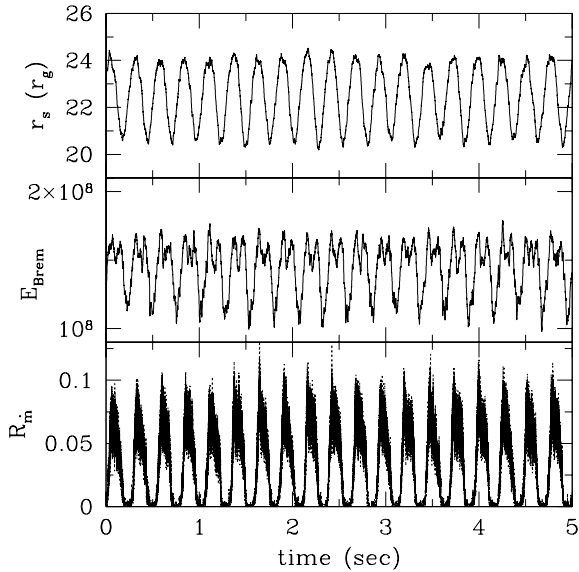}
\end{center}
\caption{{\it Top panel}: Variation of shock location with time.
{\it Middle panel}: Variation of the bremsstrahlung emission in
arbitrary units with time.
{\it Bottom panel}: Variation of mass outflow rate with time. 
Here, $\alpha = 0.011$ and $M_B = 10 M_\odot$. Other parameters are same as Figure 3.
}
\end{figure}

The quasi-periodicity that we observed in the power spectra of simulated
results seems to be generic in nature. Several Galactic black hole sources
exhibit QPO in the X-ray power spectra along with
the harmonics. In Figure 7, we plotted one such observed X-ray power spectra of black hole source
GX 339-4 of the 2010-11 outburst, which
clearly shows the presence of {\it fundamental} QPO ($\sim$ 2.42 Hz) and harmonics
at $\sim 4.88$ Hz and $\sim 7.20$ Hz \citep{ndmc12}.
This observational finding
directly supports our simulation results and 
perhaps establishes the fact that the origin of such photon flux
variation seems to be due to the hydrodynamic modulation of the
inner part of the disc in terms of shock oscillation.

Recently, \citet{nrs13}
reported the possible association of QPOs in X-rays and jets in the form of radio flares 
in outbursting black hole sources through the accretion flow dynamics.
Here also we find that the dynamics of the post-shock disc region plays a
major role for the jet generation and the emitted radiation. In other
words, post-shock disc seems to be the precursor of jets as well as QPOs 
according to our present study.

\begin{figure}
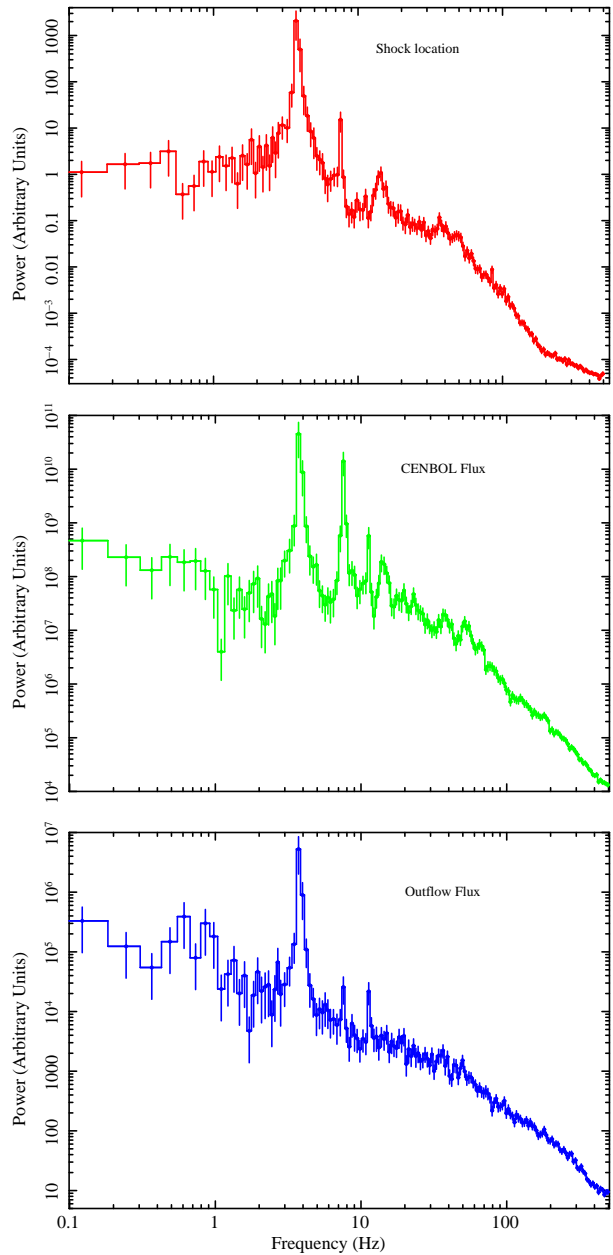

\begin{center}
\includegraphics[angle=270,width=0.45\textwidth]{fig5a.eps}
\includegraphics[angle=270,width=0.45\textwidth]{fig5b.eps}
\includegraphics[angle=270,width=0.45\textwidth]{fig5c.eps}
\end{center}
\caption{
{\it Top panel}: Fourier spectra of shock location variation at the
disc equatorial plane.
Power spectra of bremsstrahlung flux variation calculated for SPH particles
resides within the boundary of CENBOL ({\it middle panel}) and within the
outflow region ({\it bottom panel}), respectively.
Here, $\lambda_{\rm inj} = 1.63$ and $\alpha = 0.011$.
In this case, we have consider simulated data of $\sim 20$ sec. Other
parameters are same as Figure 1. Fundamental QPO frequency is obtained
in both the cases $\sim 3.7$ Hz. However, significant differences in
bremsstrahlung flux of CENBOL and outflow is observed.
}
\end{figure}
%
\begin{figure}
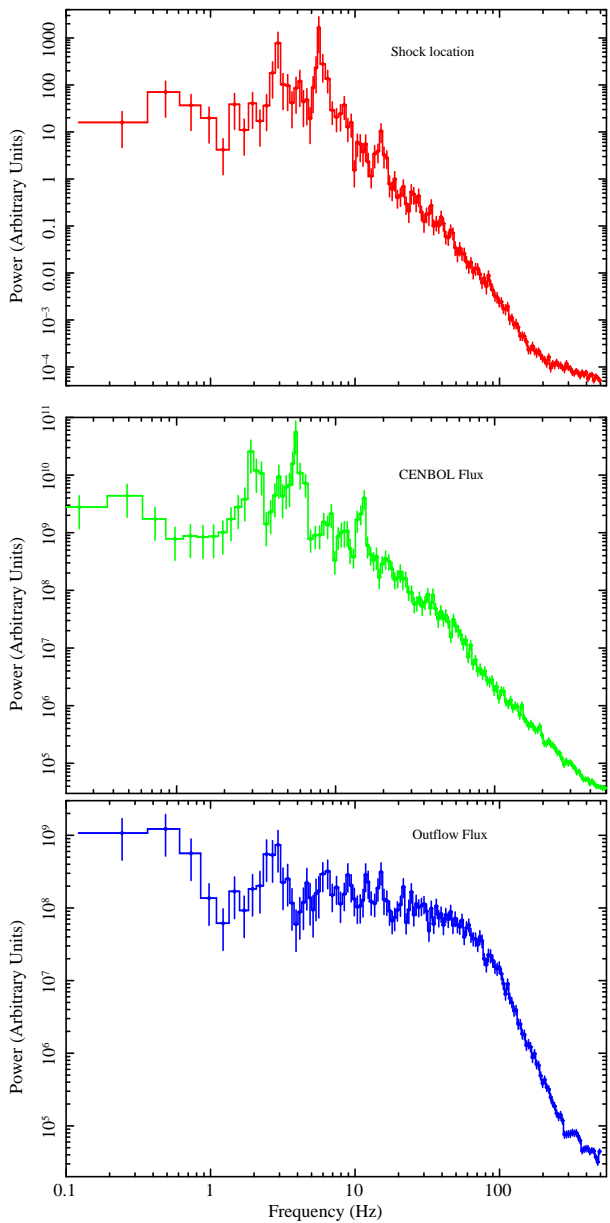

\begin{center}
\includegraphics[angle=270,width=0.45\textwidth]{fig6a.eps}
\includegraphics[angle=270,width=0.45\textwidth]{fig6b.eps}
\includegraphics[angle=270,width=0.45\textwidth]{fig6c.eps}
\end{center}
\caption{
{\it Top panel}: Fourier spectra of shock location variation at the
disc equatorial plane for $\lambda_{\rm inj} = 1.61$ and $\alpha = 0.013$. 
The other parameters are same as Figure 1.
Power spectra of bremsstrahlung flux variation calculated for SPH particles
resides within the boundary of CENBOL ({\it middle panel}) and within the
outflow region ({\it bottom panel}), respectively. 
In this case, we have consider simulated data of $\sim 15$ sec. 
Fundamental QPO frequency is obtained in both the cases $\sim 2.9$ Hz. 
}
\end{figure}

\begin{figure}
\begin{center}
\includegraphics[angle=270,width=0.45\textwidth]{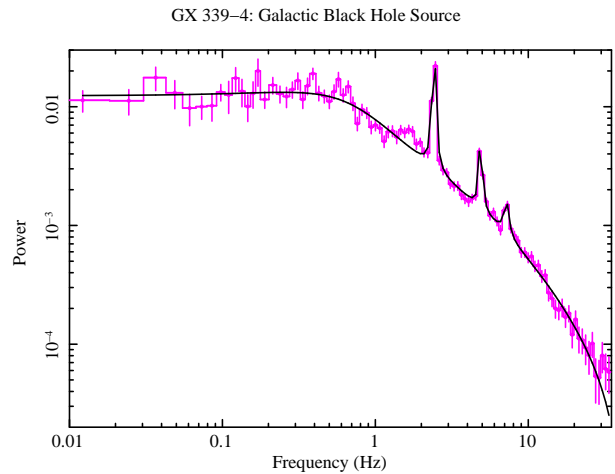}
\end{center}
\caption{
Signature of the multiple QPO frequencies ($\sim $2.42 Hz, 4.88 Hz \& 7.20 Hz)
observed in the power spectra of the galactic black hole source GX339-4.
}
\end{figure}

\section{Discussion and Concluding remarks}

We have studied the dynamics of the viscous accretion flow around 
black holes using time dependent numerical simulation. While, accreting
matter slows down against gravity as it experiences a barrier due to
centrifugal force and eventually enter in to the black hole after triggering
shock transition. Usually, post-shock flow is hot and compressed causing 
a thermal pressure gradient across the shock. As a result, it deflects
part of the accreting matter as bipolar jet in a direction perpendicular
to the disc equatorial plane. When viscosity is increased, shock
becomes non-steady and ultimately starts oscillating when the viscosity
reached its critical limit. Consequently, the outflowing matter also
starts demonstrating quasi-periodic variation. Since the inner disc is
hot and dense, high energy radiations must emit
from the vicinity of the black holes. When the inner disc vibrates
in radial direction, the emitted photon flux is also modulated.
We compute the power density spectra of such behaviour and obtain
fundamental peak at few Hz. We 
find that some of the harmonics are very prominent as seen in
the observational results of several black hole candidates.
The highlight of this paper is to show that the oscillation of shocked
accretion flow shows QPOs with fundamentals as well as harmonics (Figs. 5, 6)
as is seen from observations (Fig. 7).
Interestingly, the bipolar outflow shows at least the fundamental
frequency in its PDS for the case depicted in Figs. 5, however,
the fundamental and harmonics are fairly weak in Figs. 6.
So, this result suggests that photons from the outflows and jets would
at least show the fundamental frequency, but probably no harmonics.
Moreover, does this mean that if we happen to `see' down the length of a jet,
we would see quasi-periodic oscillations of photons in some jets (\eg Figs. 5) and in some other jets
the QPO signature would be washed out (\eg Figs. 6)? And indeed in most blazars
QPOs have not been detected, but in few QPO was found \citep{lggw09}.
This issue need further investigation. Furthermore, while hot, dissipative
flows show single shock, low energy dissipative flow
showed multiple shocks \citep{lcscbz08,lrc11}, however, the effect of high $\alpha$ has not been
investigated. 

The outflow also shows quasi-periodicity, however, blobs of matter are being ejected
persistently with the oscillation of inner part of the disc,
and therefore, such persistent activity will eventually give rise to a stream of matter and therefore a quasi-steady
mildly relativistic jet. These ejections are not the ballistic relativistic
ejections observed during the transition of hard-intermediate spectral state to the
soft-intermediate spectral state. It has been recently shown that the momentum
deposited by the disc photons on to jets, makes the jets stronger as the disc
moves from LS to hard-intermediate spectral state
\citep{kcm14}, simulations of which will be communicated elsewhere.

In this work, the mechanism studied for QPO generation is
due to the perturbations induced by viscous dissipation
and angular momentum transport. While it has been reported that QPOs can also
be generated by cooling \citep{msc96}. In realistic disc, both processes are active, and both should produce
shock oscillation. Interestingly
though, viscosity can produce multiple shocks (for one spatial dimensional results see Lee \etal 2011),
while no such thing has been reported
with cooling processes, albeit investigations with cooling processes have not been done extensively.
We would like to investigate the combined effect of cooling and viscous dissipation in future,
to ascertain the viability of `shock cascade' in much greater detail.
It must be pointed out that this model of QPO and mass ejection \citep{ncvr01} can also be applied
to the weakly magnetized accreting neutron stars. However,
one has to change the inner boundary condition, \ie put a hard surface as the inner boundary condition.
The same methodology should also give rise to QPOs, and we are working on such a scenario, and would be reported
elsewhere.

\section*{Acknowledgments}

AN acknowledges Dr. Anil Agarwal, GD, SAG, Mr. Vasantha E. DD, CDA and Dr. S. K. Shivakumar,
Director, ISAC for continuous support to carry out this research at ISAC, Bangalore. The authors
also acknowledge the anonymous referee for fruitful suggestions to improve the quality of the paper.

 \label{lastpage}

\end{document}